\documentclass[,final]{aipproc}
\layoutstyle{6x9}
\usepackage{slashed}

\newcommand{\beq}{\begin{equation}}
\newcommand{\eeq}{\end{equation}}
\newcommand{\ba}{\begin{array}}
\newcommand{\ea}{\end{array}}
\newcommand{\bea}{\begin{eqnarray}}
\newcommand{\eea}{\end{eqnarray}}

\newcommand{\ie}{{i.e.}}

\begin{document}
\title{Ginzburg-Landau approach to inhomogeneous chiral phases of QCD}

\classification{12.38.Mh, 21.65.Qr, 25.75.Nq}
\keywords      {QCD, critical point, inhomogeneous phases, chiral symmetry}

\author{Hiroaki Abuki}{
  address={Department of Physics, Tokyo University of Science,
  Tokyo 162-8601, Japan%
  },
}
\author{Katsuhiko Suzuki}{
  address={Department of Physics, Tokyo University of Science,
  Tokyo 162-8601, Japan%
  }
}

\begin{abstract}
We study the inhomogeneous chiral condensates in the proximity of the
 chiral tricritical point (TCP) of two-flavor QCD. 
Deriving the Ginzburg-Landau (GL) functional up to the eighth order in
 the order parameter and its spatial derivative, we explore off the TCP
 and find that critical curves are bent by non-linear effects.
In the newly extend GL coupling space, we find the TCP being realized
as a multicritical point where five independent critical lines meet up.
We also present general analyses for the energies associated with
 several higher dimensional crystal structures.
\end{abstract}

\maketitle

\noindent
{\bf\emph{Introduction. --}}
Chiral condensates with possible crystal structures are of particular
interest in the field of QCD under extreme conditions. 
Such a partial spatial breaking of quark-antiquark condensate may be
driven by the quark chemical potential $\mu$ 
serving
as an external
field which favor to produce a net population excess of quarks over
antiquarks. 
This was indeed proven to be true in the neighborhood of the chiral
tricritical point (TCP) in Ref.~\cite{Nickel:2009ke} where the author took the
advantage of the Ginzburg-Landau (GL) framework which is capable to make
model independent predictions.  
TCP was found to be replaced by the Lifshitz point (LP) from which
a solitonic chiral phase expands.
Such phase structure is now known to survive after incorporating the effect of current
quark mass \cite{Nickel:2009wj} or Polyakov loop \cite{Carignano:2010ac}.

Most studies so far restrict the analyses to the one dimensional
crystal structure mainly for technical reason, although such state is
known to be fragile against thermal fluctuations \cite{Baym:1982ca}.
Also moving off the TCP, we might expect other types of
inhomogeneous states, including the chiral density wave
\cite{Nakano:2004cd}, come into play and change the phase structure
just as in the case of superconductivity under an external magnetic
field \cite{Matsuo:1998mh}.

In this article, we report the extended GL analyses for the chiral
crystal phases near TCP in the chiral limit based on our recent
paper \cite{Abuki:2011pf}.
We aim at the clarification of (1)~how the inhomogeneous chiral
condensate is deformed or taken over by other phases once departing 
from the TCP, (2)~whether or not multidimensional crystal structures
get favored over one-dimensional solitonic state, and finally
(3)~the structure of the TCP in a wider extended GL coupling space.

\begin{figure}[bt]
\centering
\includegraphics[clip,width=0.48\textwidth]{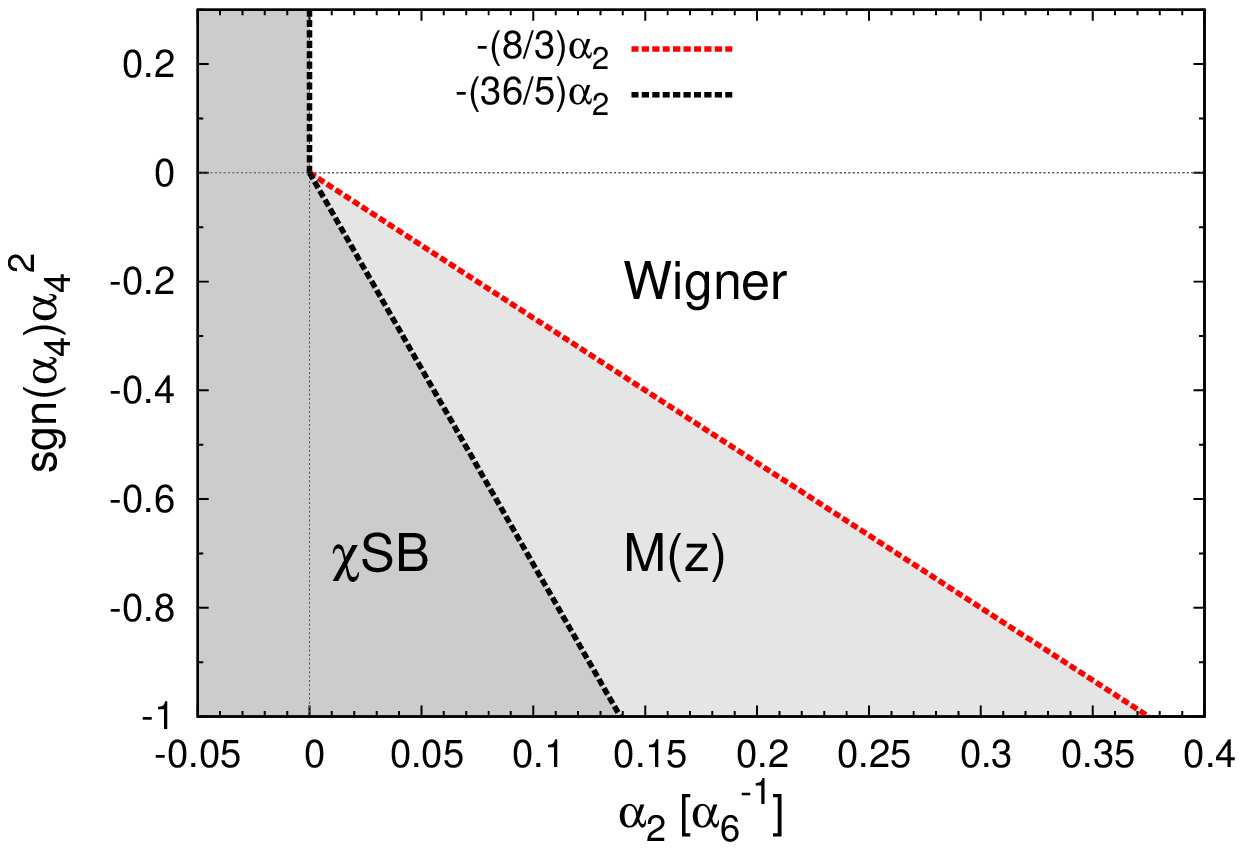}
\includegraphics[clip,width=0.48\textwidth]{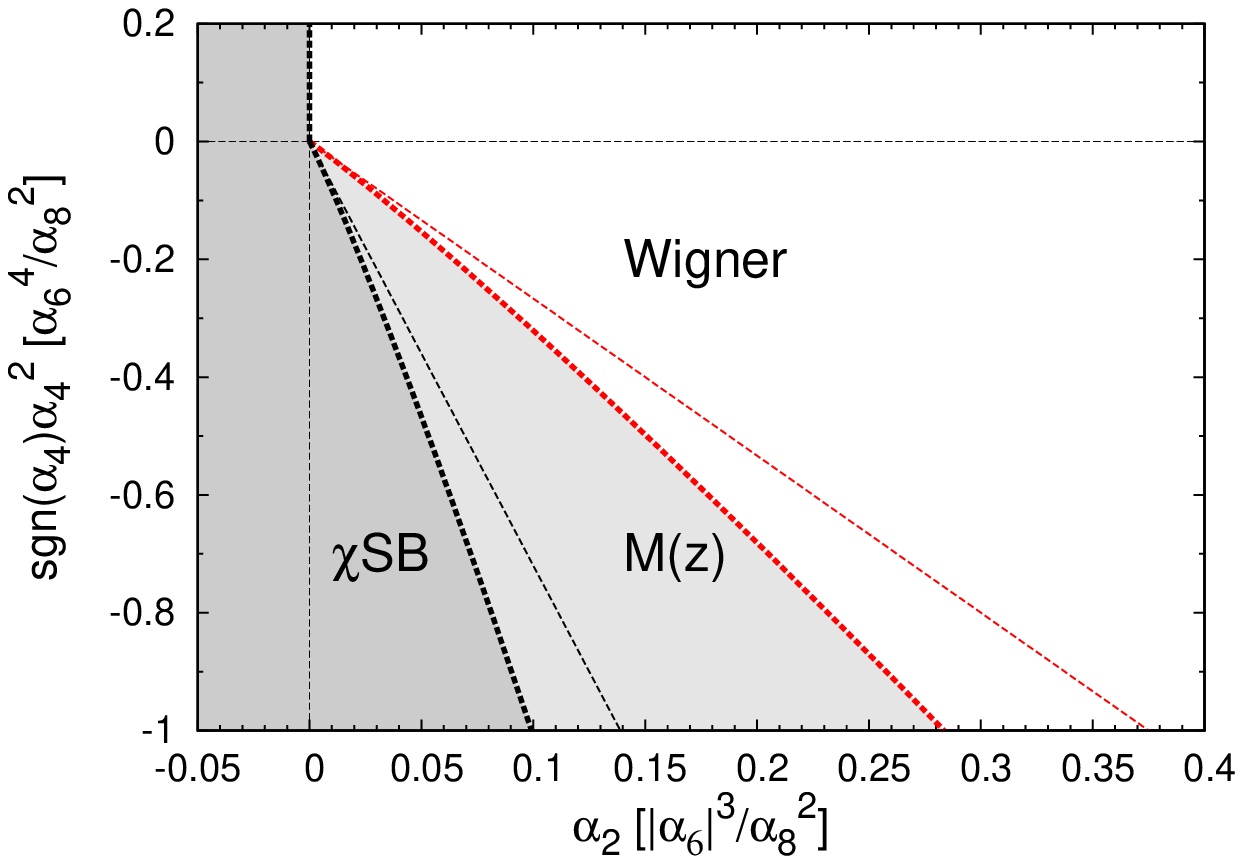}
\caption{%
(Left panel):~The GL phase diagram for Eq.~(\ref{eq:GL}), computed at
 the mean-field level with restricted to one-dimensional structures.
(Right panel):~The phase diagram after including the effect of eighth
 order terms, $\delta\omega$. For comparison, the critical lines
 enclosing the inhomogeneous phase in the absence of $\delta\omega$
 (those in the left figure) are repeated by thin dashed lines.
}
\label{fig:phase}
\end{figure}

\vspace*{1ex}
\noindent
{\bf\emph{Extended Ginzburg-Landau approach. --}}
In order to allow for a minimal description of the TCP and possible
inhomogeneous states in its neighborhood, we need to expand the GL
functional up to sixth order in the order parameter and its spatial
derivative. The expression was found as \cite{Nickel:2009ke}:
\beq
\textstyle
\omega=\frac{\alpha_2}{2}M({\textbf x})^2+\frac{\alpha_4}{4}\big(M({\textbf
x})^4+(\nabla M)^2\big)+\frac{\alpha_6}{6}\Big(M({\textbf x})^6+5M^2(\nabla
M)^2+\frac{1}{2}(\Delta M)^2\Big),
\label{eq:GL}
\eeq
where $M({\textbf x})\sim \langle \bar{q}q({\textbf x})\rangle$ is the
chiral condensate, and $\alpha_2$, $\alpha_4$ and $\alpha_6$ are
the relevant GL couplings.
If $\alpha_6>0$, We can draw a phase diagram in the two dimensional GL
parameter space $(\alpha_2,\alpha_4)$.
It can be shown that the thermodynamics only depends on the sign of
$\alpha_4$ and dimensionless ratio
$\alpha_4^2/(\alpha_2[\alpha_6^{-1}])$
where $\alpha_2[\alpha_6^{-1}]$ means $\alpha_2\alpha_6$, \ie, the value
of $\alpha_2$ measured in the unit of $\alpha_6^{-1}$. 
Due to this scaling property, all the critical lines in the
$(\alpha_2[\alpha_6^{-1}],\mathrm{sgn}(\alpha_4)\alpha_4^2)$-plane
should be expressed as straight lines, and any two of them never
intersect one another once departing from the origin. 
The phase diagram for this case is shown in the left panel of
Figure~\ref{fig:phase}. We find a Lifshitz TCP at the origin; 
the region labeled by $M(z)$ is the phase of
inhomogeneous chiral condensate characterized by Jacobi's elliptic
function $M(z)=\sqrt{k}\nu\mathrm{sn}(kz,\nu)$ with $\nu$ being the
elliptic modulus. 

\vspace*{1ex}
\noindent
{\bf\emph{Exploration off the TCP. --}}
In order to explore a new phase structure which may show up away from the
TCP within the GL framework, we need to go beyond the minimal GL
described above.
The term which should be added at eighth order is \cite{Abuki:2011pf}
\beq
\textstyle
 \delta\omega=\frac{\alpha_8}{8}\left(%
M^8+14M^4(\nabla M)^2-\frac{1}{5}(\nabla
M)^4+\frac{18}{5}M\Delta M(\nabla M)^2+\frac{14}{5}M^2(\Delta
M)^2+\frac{1}{5}(\nabla\Delta M)^2\right).
\label{eq:8th}
\eeq
Coefficients appearing in front of each term are all proportional to
$\alpha_8$, and can be extracted by evaluating the quark loop diagrams.
$\alpha_8$ should be positive for thermodynamic stability, while
$\alpha_6$ can be either positive or negative.
We first focus on the case $\alpha_6>0$ which has close connection to
the analysis at the sixth order, and later come back to the case
$\alpha_6<0$.
With inclusion of this $\delta\omega$, we can see that the
thermodynamics now is a function of two independent parameters,
$\alpha_2$ and $\alpha_4$. 
In the right panel of Figure
\ref{fig:phase}, the phase diagram is depicted
in the $(\alpha_2,\mathrm{sgn}(\alpha_4)\alpha_4^2)$-plane. In this case
two critical lines surrounding the phase are bent by the
nonlinear effect from the eight order terms. In fact it is possible to
derive the following expansions for the formation of single soliton
(the dashed curve on the left side) and for the melting of
condensate (the one on the right side, red online):
\beq
\ba{l}
\alpha_2=\frac{5}{36}\alpha_4^2%
\left(1+\frac{25}{42}\alpha_8\alpha_4+\frac{625}{784}(\alpha_8\alpha_4)^2%
+\mathcal{O}\left((\alpha_8\alpha_4)^3\right)\right),\\[2ex]
\alpha_2=\frac{3}{8}\alpha_4^2\left(1+\frac{9}{20}\alpha_8\alpha_4%
+\frac{729}{1600}(\alpha_8\alpha_4)^2+\mathcal{O}%
\left((\alpha_8\alpha_4)^3\right)\right).
\ea
\eeq
Here $\alpha_2$ and $\alpha_8$ in the above formulas mean
$\alpha_2[\alpha_6^{-1}]$ and $\alpha_8[\alpha_6^2]$ respectively.
However, apart from these nonlinear bending of critical lines, no new phase
structure appears.

\begin{figure}[tb]
\centering
\includegraphics[clip,width=0.48\textwidth]{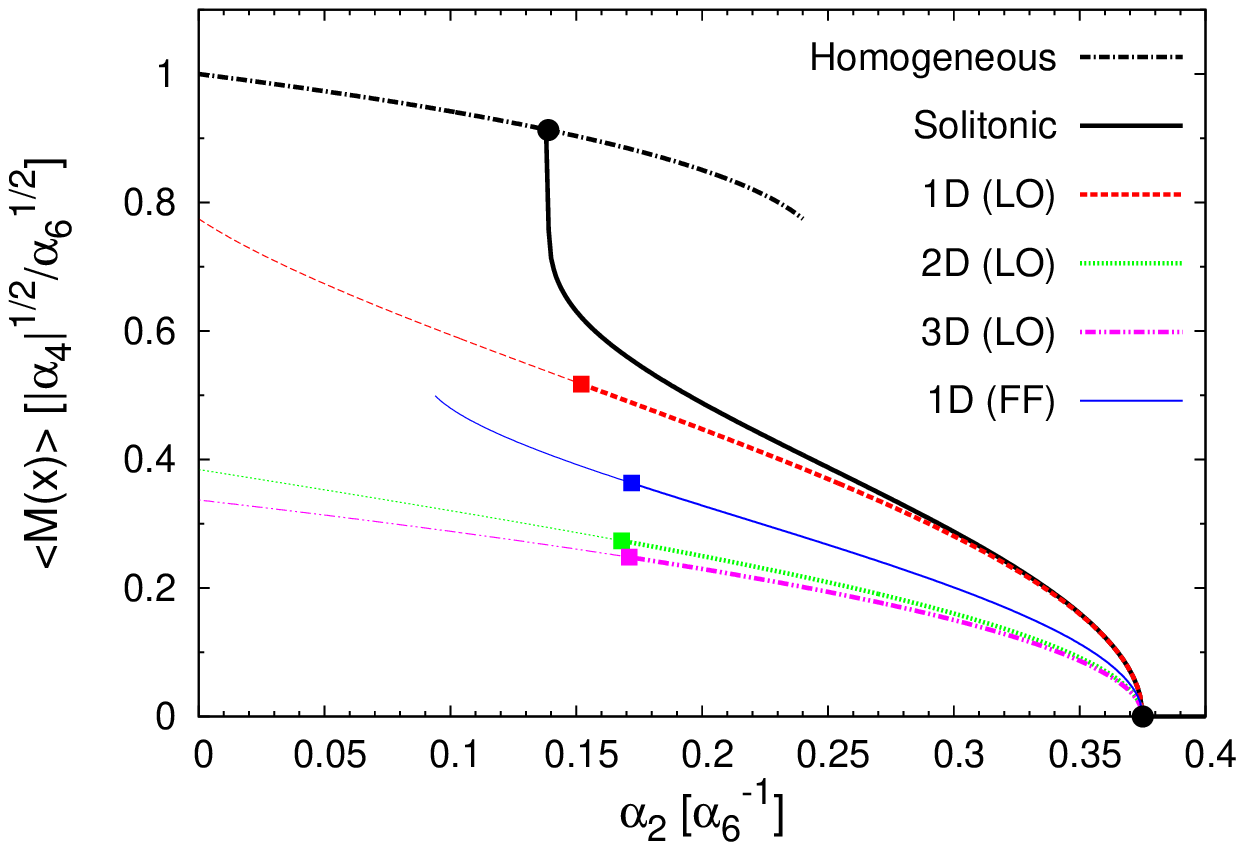}
\includegraphics[clip,width=0.48\textwidth]{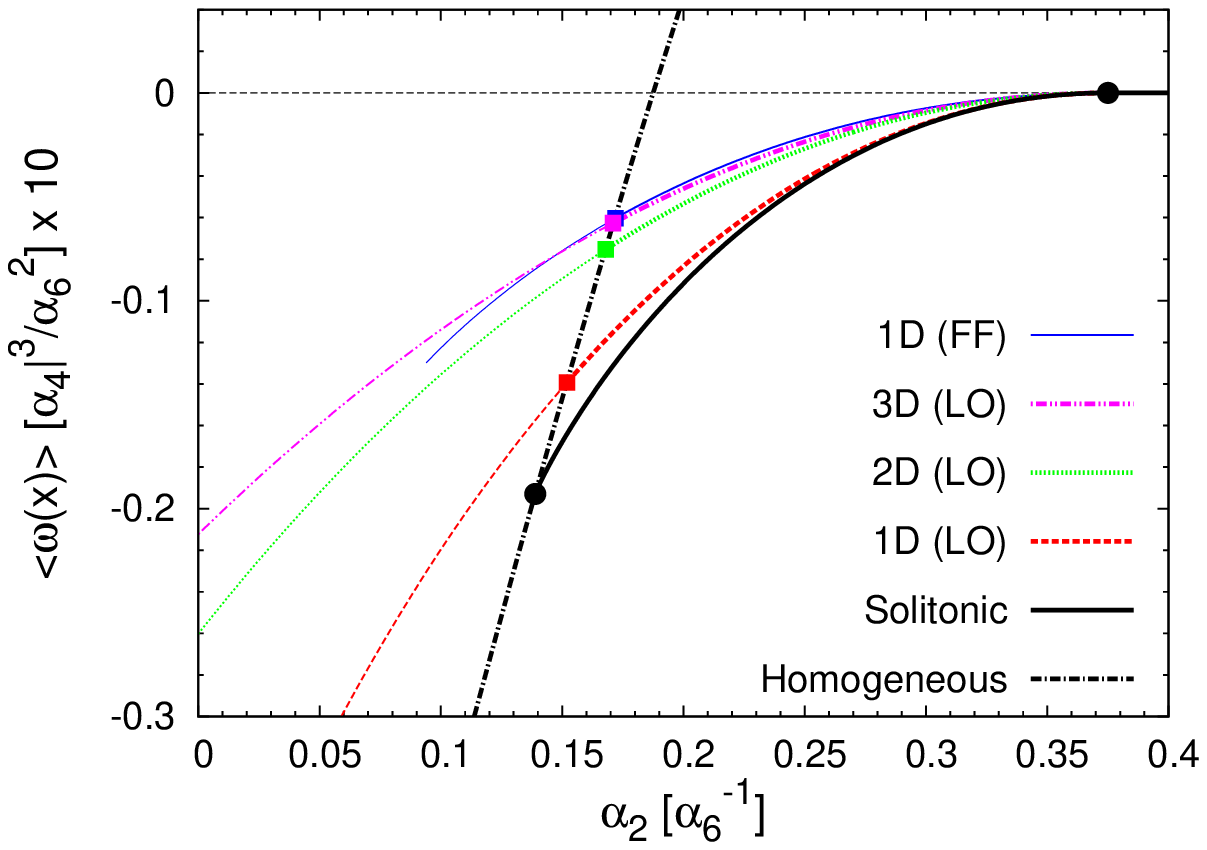}
\caption{%
The comparison of spatial average of order parameter (Left)
and that of functional potential density (Right) for several states
 along the section $\alpha_4=-1$ in the left figure of
 Figure~\ref{fig:phase}.
}
\label{fig:comp}
\end{figure}

\vspace*{1ex}
\noindent
{\bf\emph{Multidimensional crystal structures? --}}
We have limited the above
analyses to the condensate having its spatial modulation only in one
direction, \ie, one-dimensional modulation. With this
restriction, it is possible to solve the Euler-Lagrange equation
analytically leading to results demonstrated in Figure~\ref{fig:phase}.
Here we lift the limit, and explore the possibility of multidimensional
crystal structures using a variational method.
The states we consider are the extensions of one-dimensional Larkin
Ovchinnikov (LO)-type condensate
$(M_{\mathrm{1D-LO}}({\textbf x})=\sqrt{2}M_0\sin(kz))$ to higher dimensions,
\beq
\ba{l}
M_{\mathrm{2D-LO}}({\textbf x})=M_0(\sin(kx)+\sin(ky)),\\[1ex]
M_{\mathrm{3D-LO}}({\textbf
x})=\sqrt{\frac{2}{3}}M_0(\sin(kx)+\sin(ky)+\sin(kz)),
\ea
\eeq
where $k$ and $M_0$ are to be optimized for each state.
Figure~\ref{fig:comp} shows the comparison of the spatial average of
order parameter $\langle M({\textbf x})\rangle$ (Left panel) and the averaged
potential $\langle\omega({\textbf x})\rangle$ (Right panel) for the
several states along the section $\alpha_4=-1$ in the left panel of
Figure~\ref{fig:phase}. ``{FF}'' refers to the Fulde-Ferrell (FF)-like
state characterized by $M_{\mathrm{1D-FF}}({\textbf x})=M_0e^{ikz}$ in
which the time-reversal symmetry as well as rotational symmetry is
broken.
We see that the magnitude of condensate is decreasing function of
dimensionality and the energy density is accordingly in the opposite
tendency. Similar results for two dimensional condensates at $T=0$ were
recently obtained in the NJL model
\cite{Carignano:2011gr,Carignano:2012sx}.
The inclusion of eighth order terms in $\delta\omega$ does not change
this situation so that one-dimensional solitonic state is most favorable
for all values of $\alpha_8\alpha_4$.

We can derive the analytic formulas for the dependence of free energy
on the number of spatial directions in which the condensate varies
by expanding the GL functional in the vicinity of condensate-melting
line.
We first set the higher (d-)dimensional analog of the LO and FF states
as
\beq
\ba{l}
\textstyle
M_{\mathrm{dD-FF}}({\textbf x})=\frac{1}{\sqrt{d}}%
M_0(e^{ikx_1}+e^{ikx_2}+\cdots+e^{ikx_d}),\\[2ex]
M_{\mathrm{dD-LO}}({\textbf x})=\sqrt{\frac{2}{d}}%
M_0(\sin(kx_1)+\sin(kx_2)+\cdots+\sin(kx_d)).
\ea
\eeq
Then at the proximity of the line of vanishing condensate, we arrive at
the following analytic formulas for $\langle\omega({\textbf x})\rangle$ via
expanding in $M_0$ after optimizing the wavevector $k$ over $M_0$:
\beq
\ba{l}
\langle \omega_{\mathrm{dD-LO}}({\textbf x})\rangle=\left(%
\frac{\alpha_2}{2}-\frac{3}{16}\alpha_4^2\right)M_0^2+\frac{2d-1}{4d}M_0^4%
+\mathcal{O}(M_0^6),\\[2ex]
\langle \omega_{\mathrm{dD-FF}}({\textbf x})\rangle=\left(%
\frac{\alpha_2}{2}-\frac{3}{16}\alpha_4^2\right)M_0^2+\frac{1}{2}M_0^4%
-\frac{4d^2+10d-7}{12d^2}M_0^6+\mathcal{O}(M_0^8).
\ea
\eeq
From these we see that, in the case of LO-type condensates the energy
density is in the order of dimensionality of spatial modulation. In the
FF case, the difference becomes visible only at sextic order, and the
dependence on dimensionality is not monotonic; the 2D structure is most
favorable within this class of condensates.
From these analyses we find that increasing the directions of spatial
modulation is accompanied by a large kinematical energy cost, leading to
the conclusion that multidimensional crystals are unlikely to be
realized in the neighborhood of the TCP.

\begin{figure}[tp]
\centering
\includegraphics[clip,width=0.48\textwidth]{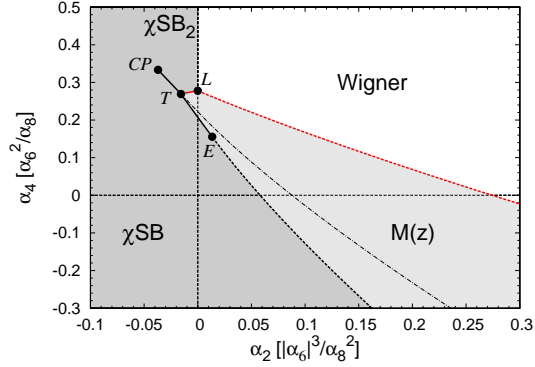}
\caption{%
The GL phase diagram for $\alpha_6<0$ computed from Eq.~(\ref{eq:GL}).
Dashed lines express the second order phase boundaries, while full lines
do the first order ones. Dot-dashed line in the inhomogeneous phase
 represents the first order phase transition which would have been
 realized if the inhomogeneous phase was not taken into consideration.
}
\label{fig:phase2}
\end{figure}

\vspace*{1ex}
\noindent
{\bf\emph{The structure of TCP. --}}
We now explore a new regime $\alpha_6<0$. The corresponding
phase diagram is depicted in Figure~\ref{fig:phase2}.
We see that the Lifshitz tricritical point for $\alpha_6>0$ splits into
four multicritical points; the critical point
({\bf\emph{CP}}), the Triple point ({\bf\emph{T}}), the Lifshitz
bicritical point ({\bf\emph{L}}), and the critical endpoint
({\bf\emph{E}}).
Their $(\alpha_2,\alpha_4)$-coordinates are listed in TABLE~\ref{tab:1}.
All these points scale as
$(\#|\alpha_6|^3/\alpha_8^2,\#\alpha_6^2/\alpha_8)$ so that they are 
all going to degenerate into the origin as $\alpha_6\to0^-$, smoothly
continuing to the Lifshitz TCP in the positive $\alpha_6$
side.
Among these critical points the most intriguing is {\bf\emph{T}} where
three first order phase transitions meet. At this point, three different
forms of broken-symmetry phase compete and coexist. These are; the state
with homogeneous chiral condensate, the state with smaller chiral
condensate denoted by ``$\chi$SB$_2$'', and the inhomogeneous chiral
condensate labeled by $M(z)$.
It is possible to extract numerically the dimensionless ratios of
physical quantities that are to be realized when the point 
{\bf\emph{T}} is
approached:
\beq
\lim_{(\alpha_2,\alpha_4)\to\tiny T}\bigg\{%
\frac{M(\chi\mathrm{SB}_2)}%
{M(\chi\mathrm{SB})},%
\frac{\sqrt{\langle
M(z)^2\rangle}}{M(\chi\mathrm{SB})},%
\frac{2\pi/\ell_\mathrm{P}}%
{\sqrt{\langle M(z)^2\rangle}}
\bigg\}=\left\{0.369,~0.308,~5.001\right\},
\eeq
where $\ell_\mathrm{P}$ is the modulation period.
These ratios characterize the triple point ({\bf\emph{T}}) and they are
universal and model-independent in the sense that they do not depend on
any of GL parameters including the magnitude of $|\alpha_6|$, and
$\alpha_8$. 

\begin{table}[tp]
\caption{The coordinates of locations of the multicritical points.}
\begin{tabular}{rlll}
\hline
Label & \multicolumn{1}{l}{Type} & \multicolumn{2}{l}{%
Coordinate $(\alpha_4~[\alpha_6^3/\alpha_8^2],%
\alpha_2~[\alpha_6^2/\alpha_8])$}\\ \hline
{\bf\emph{CP}} & Critical point & $(-1/27,1/3)$ &(analytical)\\
{\bf\emph{T}} & Triple point & $(-0.016,0.27)$ &(numerical)\\
{\bf\emph{L}} & Lifshitz bicritical point & $(0,5/18)$ &(analytical)\\
{\bf\emph{E}} & Critical endpoint & $(0.014,0.16)$ &(numerical)\\ \hline
\end{tabular}
\label{tab:1}
\end{table}

\vspace*{1ex}
\noindent
{\bf\emph{Summary. --}}
In conclusion, we explored off the TCP based on the GL functional
expanded up to eighth order in chiral order parameter and its spatial
derivative. As a consequence, we found that the critical lines
surrounding the inhomogeneous phase are bent because of the nonlinear
effect coming from eight order terms although the qualitative phase
structure remains unaffected. We also examined the possibility of higher
dimensional crystal phase, but it turned out that the kinematical cost
to make the condensate modulated in several directions is so high that
it is unlikely to have multidimensional chiral crystals near the TCP. 
Finally we extended our analyses to a new regime $\alpha_6<0$. We found
that the TCP splits into four multicritical points. It would be interesting
to search for a thermodynamic physical quantity in QCD to which
$\alpha_6$ is sensitive. The isospin chemical potential or the chiral
chemical potential might serve as one of such sources. The
investigations in these directions deserve further efforts in future
\cite{Iwata:2012bs}. 

\vspace*{1ex}
\noindent
{\bf\emph{Acknowledgments. --}}
H. A. thanks the organizers of QCD@Work2012 for giving him an
opportunity to give a talk.
The authors thank D.~Ishibashi for the fruitful collaboration.

\bibliographystyle{aipproc}   

\begin{thebibliography}{99}

\bibitem{Nickel:2009ke} 
  D.~Nickel,
  Phys.\ Rev.\ Lett.\  {\bf 103}, 072301 (2009)
  [arXiv:0902.1778 [hep-ph]].

\bibitem{Nickel:2009wj} 
  D.~Nickel,
  Phys.\ Rev.\ D {\bf 80}, 074025 (2009)
  [arXiv:0906.5295 [hep-ph]].


\bibitem{Carignano:2010ac} 
  S.~Carignano, D.~Nickel and M.~Buballa,
  Phys.\ Rev.\ D {\bf 82}, 054009 (2010)
  [arXiv:1007.1397 [hep-ph]].

\bibitem{Baym:1982ca} 
  G.~Baym, B.~L.~Friman and G.~Grinstein,
  Nucl.\ Phys.\ B {\bf 210}, 193 (1982).

\bibitem{Matsuo:1998mh}
  S.~Matsuo, S.~Higashitani, Y.~Nagato, and K.~Nagai, 
  J.\ Phys.\ Soc.\ Jpn.\ {\bf 67}, 280 (1998).

\bibitem{Abuki:2011pf} 
  H.~Abuki, D.~Ishibashi and K.~Suzuki,
  Phys.\ Rev.\ D {\bf 85}, 074002 (2012)
  [arXiv:1109.1615 [hep-ph]].

\bibitem{Nakano:2004cd} 
  E.~Nakano and T.~Tatsumi,
  Phys.\ Rev.\ D {\bf 71}, 114006 (2005)
  [hep-ph/0411350].

\bibitem{Carignano:2011gr} 
  S.~Carignano and M.~Buballa,
  arXiv:1111.4400 [hep-ph].

\bibitem{Carignano:2012sx} 
  S.~Carignano and M.~Buballa,
  arXiv:1203.5343 [hep-ph].

\bibitem{Iwata:2012bs} 
  Y.~Iwata, H.~Abuki and K.~Suzuki,
  arXiv:1206.2870 [hep-ph].

\end{thebibliography}

\end{document}